\documentclass{article}

\input{epsf}

\begin{document}
\fontsize{12}{14}\selectfont

\title{Glass transition in self organizing cellular patterns}   

\author{ {Tomaso Aste* and David Sherrington}  \\
{\it	Dep. of Physics, Oxford University, Theoretical Physics, }\\
{\it	 1 Keble R., Oxford OX1 3NP, England } \\
{\it	{*} INFM, C.so Perrone 24, Genova Italy and }\\
{\it	LDFC,  3 rue de l'Univresit\'e, Strasbourg France.} \\
{\it 	aste@infm.it }
	}

\date{\today}

\maketitle

\section*{Abstract} 

{\it
We have considered the dynamical evolution of cellular
patterns controlled by a stochastic Glauber process
determined by the deviations of local cell topology
from that of a crystalline structure.  Above a critical
temperature evolution is towards a
common equilibrium state from any initial configuration,
but beneath this temperature there is a dynamical phase
transition, with a start from a quasi-random state leading
to non-equilibrium glassy freezing whereas an ordered
start rests almost unchanged.
A temporal persistence function decays exponentially
in the high temperature equilibrating state but has a
characteristic slow non-equilibrium aging-like behaviour in the low
temperature glassy phase.
}

\bigskip
The hexagonal tiling is the best partition of the plane in equal cells.
It solves both the packing and the covering problems.
It is the space-filling assembly of equal  cells with the minimal
interfacial extension.

It is regular, perfect and beautiful.
Surprisingly, however, it is never realized in natural biological
tissues where a
relevant amount of disorder is always present \cite{Lew,WR84}.
This is reminiscent of the situation found in many covalently
bonded solids, where, despite the lowest energy state being crystalline,
in practice amorphous glassy structures are the common quasi-stable
states \cite{Zach}.   Guided by this observation and by recent
analysis of idealized infinite-ranged spin glasses\cite{bouchaud}
and in a desire to provide both a linkage to studies of covalent
glasses and a minimalist model for understanding, we have studied
the dynamics of the simplest `covalently bonded' network subject
to a simple stochastic dynamics characterized by a minimal set of
control parameters, namely (i) deviations of local topology from
crystalline and (ii) a temperature.  The model has similarities
with those of the classic works of Weber et. al.\cite{DS1}
and Wooten et. al.\cite{DS2,DS3}, but complements and extends
those studies by emphasising (and quantifying) not only the
freezing of single time measures and slow relaxation, but also
the apparent non-equilibrating non-stationary (aging)
character of two-time correlation observables in the glassy
phase.  It therefore provides a link between the two fields
of glass studies as epitomized by reference\cite{bouchaud}
on the one hand and references\cite{DS1,DS2} on the other.
It should provide a valuable starting point for further
theoretical analysis of systems complementary to those studied
in\cite{bouchaud} and in recent work on glass transitions in
model systems with central forces\cite{DS4}.  It is also
extendable, at least in principle, to higher
dimensions/more-armed vertices.  Our study combines
Monte Carlo simulation and approximate analysis and concentrates
on simple but novel one-time and two-time observables, introduced
to probe issues of the character of those emphasised in\cite{bouchaud}.

In the hexagonal tiling each cell has six neighbours

and  three cells meet at a common vertex.
It is a regular  two-dimensional, minimally-connected pattern.
More generally we can consider disordered cellular patterns with
3-connected vertices such that the average number of sides per cell
is 6 (Euler's theorem \cite{WR84}).
Locally, the degree of homogeneity can be measured in terms
of the number of sides per cell.
If $ n_i$ is the number of sides of cell  $\; i$, then $q_i=6-n_i$ is a
measure of its deviation from the hexagonal configuration.
This quantity is the topological charge of the cell and is related to  the
local curvature.
Local vertex re-arrangements conserve the total topological charge
but allow migration and the annihilation of opposite charges on
adjacent cells.  When penalties are imposed for non-zero charges
this mechanism leads the system to self organize.

\begin{figure}
\vspace{-3cm}
\epsfxsize=12.cm
\epsffile{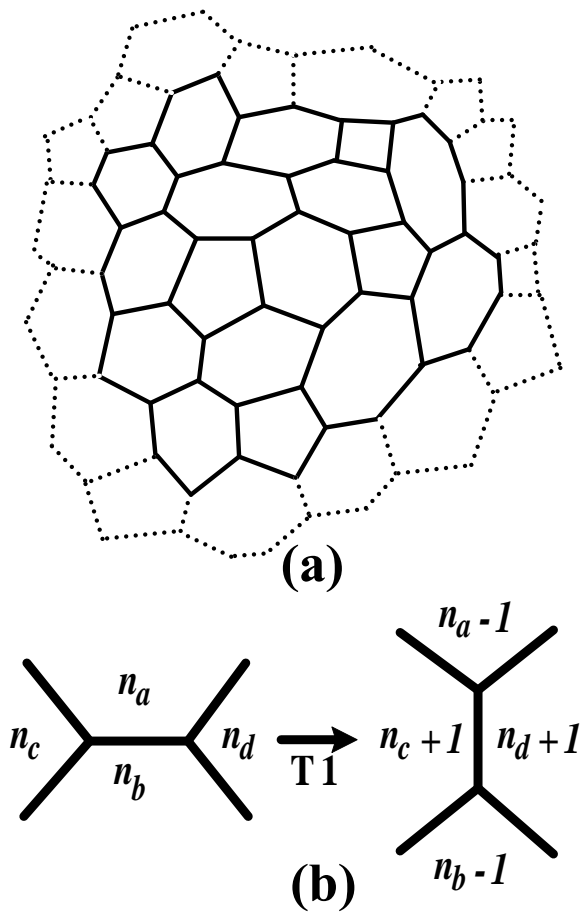}
\begin{centering}
\vspace{-1cm}
\caption{\label{f.T1B} {\bf (a)} A topologically stable cellular partition
(froth). {\bf (b)} A  T1 move.  }
\end{centering}
\end{figure}

Let us now turn to quantification.  We characterize any structure by an
`energy' and consider random sequential stochastic dynamics
driven by changes in that energy.  The energy is a measure of the
inhomogeneity in the structure.  In the spirit of minimalism we
take it to be given by the `topological
distance' from the hexagonal lattice configuration:
\begin{equation}
E = \mu_2 N = \sum^N_{i=1} (q_i)^2 = \sum_{i=1}^N (6 - n_i)^2.
\nonumber
\label{energy}
\end{equation}
We consider a system conserving cell number ($N$).  It evolves  by
T1  moves,  which are topological transformations
consisting of an  exchange of neighbours
between 4 cells [see fig.(\ref{f.T1B})].
Consider a T1 move performed on a system of four cells where the two
adjacent cells have $n_a$ and $n_b$ sides
and the two second neighbour cells have $n_c$ and $n_d$ sides respectively.
After the T1 move, the cells with $n_a$ and $n_b$ sides lose one edge each,
whereas the cells with $n_c$ and $n_d$ sides each gain an edge.  The change
of the energy associated with this move is
$$
\Delta E(n_a,n_b;n_c,n_d)= 2 (2 + n_c + n_d - n_a - n_b )
$$
We consider dynamics of a  Glauber-Kawasaki type, where the probability of
this T1 move is given by
\begin{eqnarray}
& & \Pi(n_a,n_b;n_c,n_d)= {1 \over 1+
\exp(\beta \Delta E(n_a,n_b;n_c,n_d))}\nonumber \\
& & \ \ \ \ \times (1-\delta_{n_a,3})(1-\delta_{n_b,3})(1-\delta_{c,d})
\label{P}
\end{eqnarray}
This probability  allows a dynamical evolution even at zero temperature
($\beta = \infty$) if it reduces the energy or leaves it  unchanged.
In Eq.(\ref{P}) the first two $\delta$-function terms have been
introduced to exclude moves that  generate two-sided cells.
The last term forbids moves which generate cells which are self
neighbours, preventing the formation of tadpoles.
Thus neither two-sided nor single-sided cells can  arise if not present
at the start.

Before considering the dynamics, let us consider the predictions of
equilibrium.
The partition function is the sum over all the possible froths with
Boltzmann weights.
Some limitations apply  on the accessible values of the $n_i$.

For instance, the average number of neighbours ($\langle n \rangle$) is
fixed to be equal to 6 \cite{WR84}.
Moreover, two-sided cells are not admitted (therefore  $n_i > 2$).
Self-neighbouring cells are also not admitted (which  trivially implies $n_i
< N $),
but also other less trivial combinations of $n_i$ might be forbidden.
In general, to find the sets of admissible $\{ n_i \}$ which lead  to
possible froths is a very difficult task.

We approximate the partition function:
\begin{eqnarray}
Z(\beta,\lambda,N) \simeq  \prod_{j=1}^N \sum_{n_j=3}^{N-1}
\exp[-\beta \sum_i (6-n_i)^2 - \lambda \sum_i  (6 - n_i)],\nonumber
\end{eqnarray}
where $\lambda$ is a Lagrange multiplier fixed by the constraint $\langle n
\rangle =6$.
Correspondingly, the probability of an $n$-sided cell in the system is
\begin{equation}
p(n)\simeq p(6)\exp[-\beta(6-n)^2 - \lambda(6 - n)].
\label{pn}
\end{equation}
When $N \rightarrow \infty$ and $\beta =0 $, $\lambda = \ln(3/4)$ and $p(n)
= {16 / 27} ( {3 / 4} )^n $,
with $\mu_2 = 12$.

\begin{figure}
\vspace{-1cm}
\epsfxsize=12.cm
\epsffile{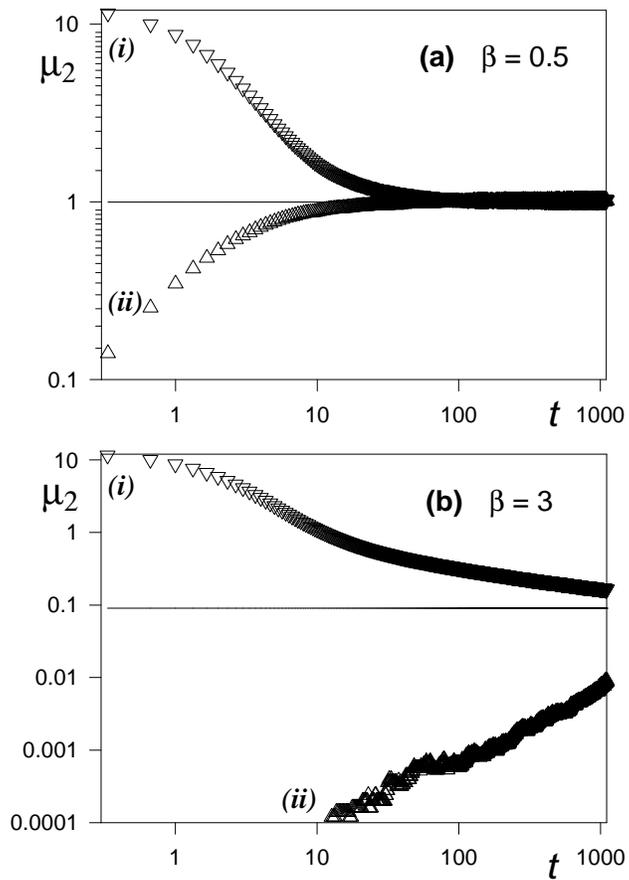}
\begin{centering}
\vspace{-2cm}
\caption{\label{f.simul}
{\bf (a)} $\mu_2$ versus the number of attempted moves
in units of  $N$  at $\beta =0.5$ with starting configurations
(i) and (ii) (triangles down and up).
\label{f.mu2glass} {\bf (b)} $\beta = 3$.
The lines are the theoretical predictions for thermodynamical equilibrium.}
\end{centering}
\end{figure}

To study the dynamics we performed extensive computer simulations on froths
with $N= 100172$ ($=317^2$) cells and periodic boundary conditions.
We  started  from two different systems:
{\bf (i)} a very disordered network, obtained by performing $10^4N$ T1 moves
on edges chosen at random from an ordered hexagonal seed;
{\bf (ii)} a perfectly ordered pattern (the hexagonal tiling).
In the simulation, $ 1100 N$  T1 moves were attempted, with a
probability given by  Eq.(\ref{P}),
on edges chosen at random.  First we consider a moderate-high temperature.
In fig.(2a) $\mu_2$ versus the number of attempted moves when $\beta =0.5$
is shown for the two starting froths.
These systems evolve from the two initial states (i) and (ii) with $\mu_2
\simeq 13$ and $\mu_2 = 0$ \cite{comment1} respectively,
toward a final equilibrium configuration with a common  $\mu_2 \simeq 1 $
reached after about $100 N$ attempted moves.
Second we turn to a lower temperature.  In fig.(2b) $\mu_2$  is plotted
versus the number of attempted moves for  $\beta =3$.
In this case, after $1100N$ attempted moves, the equilibrium configuration
is not yet reached, the two patterns generated from the two initial states
(i) and (ii) are still statistically different,
and the approach toward a common equilibrium configuration is very slow.

\begin{figure}
\vspace{-1cm}
\epsfxsize=12.cm
\epsffile{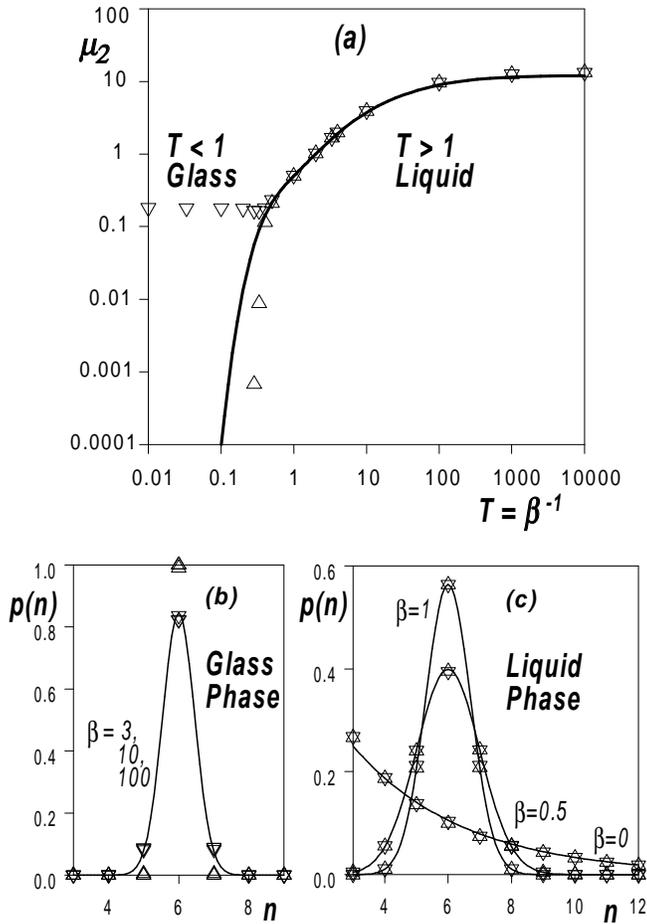}
\begin{centering}
\vspace{-2cm}
\caption{\label{f.mu2.phas.diag} {\bf (a)}$\mu_2$ versus
temperature ($T=\beta^{-1}$) with starting configurations (i) and (ii)
(triangles down and up)
dynamically equilibrated for $1100 N$  moves.
The full curve is the theoretical prediction at the
thermodynamical equilibrium.
{\bf (b)}  Side distributions ($p(n)$) at low temperatures in the ``glass
phase''
($\beta=3, 10 ,100)$.
The curve is the equilibrium prediction for $\beta=2.4$. {\bf (c)} Side
distributions  at high temperatures in the ``liquid phase''
($\beta=0,0.5,1)$.
The full lines are theoretical equilibrium predictions. }.
\end{centering}
\end{figure}

In fig.(\ref{f.mu2.phas.diag}) $p(n)$ and  $\mu_2$ are plotted for several
simulations performed at different temperatures,
starting from the two initial states (i) and (ii) and attempting $1100 N$
moves.
At high temperatures ($\beta^{-1} > 1$) the two final distributions $p(n)$
and  $\mu_2$ corresponding to the starting configurations (i) and (ii)

coincide  and are in good agreement with the analytical prediction
\cite{comment1}.
The system has reached thermodynamical equilibrium and we designate
the state as a `liquid'.
At low temperatures ($\beta^{-1} < 1$) the $p(n)$ and $\mu_2$ reached
do not coincide any more.
The dynamics is slow and the system no longer reaches  the thermodynamical
equilibrium state in the time-scale studied.
In the low temperature region for $T < O(1)$, an ordered start
rests almost unchanged and a random start seems to lead to
a freezing with $p(n)$ close to the form given by Eq.(\ref{pn}) with $\beta =
2.4$.
Hence we identify $T_d \simeq (2.4)^{-1}$ as the dynamical freezing
transition
for the time-scale studied and we designate the corresponding
metastable phase as a glass.

\begin{figure}
\vspace{-2cm}
\epsfxsize=12.cm
\epsffile{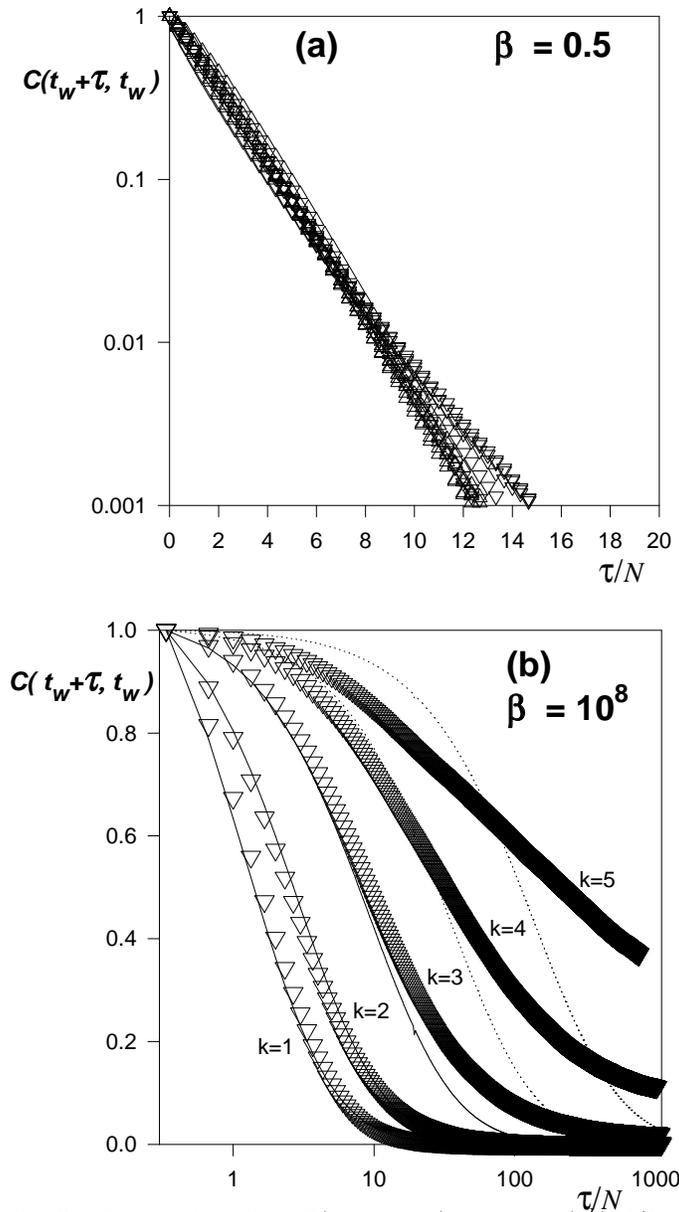}
\begin{centering}
\vspace{-1cm}
\caption{\label{f.ac.low} Persistence  function $C(t_w+\tau,t_w)$ for
$t_w=(4^k)N/3$ with $k=1,..,5$.
 {\bf (a)} At high temperatures ($\beta=0.5$) the persistence function is
independent of $t_w$
and decreases exponentially fast.
{\bf (b)} At low temperatures ($\beta=10^8$) $C(t_w+\tau,t_w)$ depends on
$t_w$ and
shows a slow decay.   The curves correspond to fits of the form
$\overline{C(t_w+\tau,t_w)} \sim C_0 ({t_w +t_0 \over  t_w+ t_0 +
\tau})^\alpha$ with $\alpha$ = 2.5, $t_0=2N$ and $C_0=N$.}
\end{centering}
\end{figure}

In order to better characterize these two different dynamics at  high and
low temperatures,  we have also studied a
{\it two time persistence function}  $C(t_w+\tau,t_w)$ \cite{Derr94}.
It counts the fraction of cells that have not been involved in a T1 move
between the time $t_w$ and $t_w + \tau$.
In fig.(\ref{f.ac.low}) the persistence functions  for $\beta = 0.5$ and $10^8$,
are  reported versus $\tau$ for several value of $t_w$.
At high temperatures the persistence function appears time-translationally
invariant and decays exponentially fast as in a {\it liquid},
whereas at  low temperatures  it shows  a  slow aging-like behaviour as in a
{\it glass} \cite{Struik,bouchaud}.  Thus, at low temperatures the system has
not equilibrated, even though the single-time measures are no longer
evolving significantly.

Let us now turn to theoretical expectations for $C$ and consider a given cell
$i$.
The probability that between times $t$ and $t+1$ one of its edges  or one of
the edges incident on its vertices  ($2n_i$ edges in total)
is chosen  (among the $3N$ edges in the system) to attempt a T1 move is $2
n_i(t)/3N$.
This T1 move is effectively performed with a probability $A_i(t)$ that
depends on the local configuration
as in  Eq.(\ref{P}).
The probability that a given cell $i$ is not involved in any move between
the time $t_w$ and $t_w+\tau$ is therefore
\begin{eqnarray}
& &\overline{ c_i(t_w+\tau,t_w)  }=  \prod_{t=t_w}^{t_w+\tau}[1-{2 n_i(t)
\over
3N} A_i(t) ] \nonumber \\
& &\sim  \exp[- \sum_{t=t_w}^{t_w+\tau}{2 n_i(t) \over 3N} A_i(t) ] \;\;\;,
\label{A3}
\end{eqnarray}
and the expectation value of the persistence is
$\overline{ C(t_w+\tau,t_w)}  = \sum_i \overline{ c_i(t_w+\tau,t_w) }$.

At high temperatures in thermodynamical equilibrium we expect that the local
configuration around the
cell $i$ evolves through a set of similar configurations typical of the
whole system.
Therefore, for large $\tau$, $\sum_{t=t_w}^{t_w+\tau} 2  n_i(t) A_i(t)
\simeq  2 \overline{n A} \tau$.
This leads to a persistence function which is independent of $t_w$ and
decreases exponentially fast.
This prediction is in good agreement with the results from simulations for
$\beta \le 1$ where we obtain best-fit  values for $\overline{n A}$ equal to
0.9, 0.83 and 0.42 for $ \beta$= 0, 0.5 and 1 respectively.

At low temperatures, almost only 5, 6 and 7 sided cells are present in the
system.
We can restrict consideration to T1 moves which either leave the energy
unchanged or reduce it.  The former act to move pairs of 7-sided cells,
pairs of 5-sided cells and 5-7 couples with adjacent pairs of 6-sided cells.

Pairs of 5-7 couples are brought together by this motion and
annihilate to produce four

6-sided cells.
The probability that between time $t$ and $t+1$ a 5-7 couple produces a T1
move in a given cell $i$ is proportional to
the number $M_{5|7}$ of 5-7 couples in the system: $ \overline{n A} \propto
M_{5|7}$.

The cellular pattern evolves reducing the number of 5-7 couples.
This happens when  two 5-7 couples are close together and a T1 move

annihilates both.
The number of couples that annihilate per unit of time (${d M_{5|7} \over
dt}$)  is proportional to the probability for two
couples to be close  together which, ignoring correlations, is
proportional to $({M_{5|7} \over N})^2 $.
This implies $M_{5|7} \propto {(t + t_0)^{-1}}$ and consequently
$\sum_{t=t_w}^{t_w+\tau} 2  n_i(t) A_i(t) \sim \alpha  \ln(t_w +t_0+ \tau)
-\alpha \ln(t_w+t_0)$, with $\alpha$ a coefficient which  depends on the
details of the annihilation process.
By substituting into Eq.(\ref{A3}), we get
$ \overline{C(t_w+\tau,t_w)} \sim C_0 ({t_w + t_0 \over  t_w + t_0 +
\tau})^\alpha$.
In fig.(\ref{f.ac.low}) this expression is compared with the results  from
the simulations for $\beta = 10^8$.
When $t_w < 10 N$ the behaviour of the persistence function is in good
qualitative agreement with this crude theoretical prediction  with fit
parameters $\alpha \simeq 2.5$, $t_0 = 2N$ and  $C_0 =1$, but  for large
waiting times $t_w > 100N$  the dynamics become slower.

In fact the recent spin-glass work\cite{bouchaud} emphasises that
aging in two-time quantities continues for both waiting time $t_w$
and relative time $\tau$ much greater than the characteristic
settling time-scale for one-time quantities.  Hence it would be
of interest to extend these studies to much longer run-times, as
well as to higher dimensions.

In summary, in a minimalist model of a continuous network
subject to stochastic dynamics determined by a simple
local topological `energy' and a randomizing temperature
we have identified a dynamical phase transition and
have  distinguished  two phases: {``\it liquid''} at high temperatures and
{\it ``glass''}  at low temperatures.
The  dynamics is {\it fast} and equilibrating from any starting
state in the {\it liquid}  phase and {\it slow} and {\it aging} in the
{\it glass} phase which is achieved from a random starting state.
In our explicit characterization
the dynamics become slow for $\beta < 1$ and the transition to the
glass phase is around the point $\beta =2.4$.

Although the micro-dynamics we have considered is idealized
and chosen for minimalist study, it is interesting to
compare our results with those observed in nature.  Most
undifferentiated biological  tissues are space-filling assemblies
of  cells  where  the  side distribution
is centered  around $n=6$ and  the  second moment  $\mu_2$ takes values
between   0.5 and 1.2; for instance, $\mu_2 = 0.53$ in {\it human epidermis
} \cite{Dub97},  $0.6 \le \mu_2  \le 1.1$  in vegetable leaves
\cite{Momba90}, $\mu_2 = 0.68 $ in  {\it cucumber epithelia}  and
$\mu_2 = 1.00 $ in {\it human amnion } \cite{Lew}.

We have found   slow  dynamics   in systems with   low
amount of disorder ($\beta >1$ and $\mu_2 < 0.5$) and
fast  dynamics  in  systems with higher disorder
($\beta <1$ and $\mu_2 > 0.5$).
One might speculate that the `ideal' biological tissue must fit
the compromise  between low disorder (homogeneity) and

fast  dynamics (efficient recovering of perturbations),
implying a structure with a value of $\mu_2$ a little above 0.5.

\bigskip

{\bf Acknowledgements }

\nobreak
\noindent
T. Aste acknowledges partial support from the European Council  (TMR
contract  ERBFMBICT950380).
A special thanks to Helgo Ohlenbuch who wrote the original C code which has
been modified and used for the present simulations.

\end{document}